\documentclass[conference]{IEEEtran}
\makeatletter
\def\footnoterule{\relax%
  \kern-5pt
  \hbox to \columnwidth{\hfill\vrule width 0.8\columnwidth height 0.4pt\hfill}
  \kern4.6pt}
\makeatother
\usepackage{xcolor}
\usepackage{extarrows}
\usepackage{stackengine}
\usepackage{tikz}
\usepackage{listings}
\usepackage{amsmath,amsfonts}
\usepackage{graphicx}
\usepackage{textcomp}
\usepackage{soul}
\usepackage{wrapfig}
\usepackage{multirow}
\usepackage{cite}
\usepackage{hyperref}
\hypersetup{colorlinks=true, unicode=true, linkcolor=[rgb]{0.10,0.05,0.67}, citecolor=[rgb]{0.10,0.05,0.67}, filecolor=[rgb]{0.10,0.05,0.67}, urlcolor=[rgb]{0.10,0.05,0.67}}
\usepackage{algpseudocode}
\usepackage{colortbl}
\usepackage{listings}

\begin{document}

 \title{\vspace*{-0.0em}Honest to a Fault: Root-Causing Fault Attacks with Pre-Silicon RISC Pipeline Characterization}
 \vspace{-1.5em}


\author{

\IEEEauthorblockN{Arsalan Ali Malik, Harshvadan Mihir, and Aydin Aysu 
\IEEEauthorblockA{Department of Electrical and Computer Engineering\\
North Carolina State University\\
Raleigh, North Carolina, 27601\\
\{aamalik3, hmihir, aaysu\}@ncsu.edu\\
\vspace*{-4.25em}
}
}

}

\maketitle
\thispagestyle{plain}
\pagestyle{plain}

\begin{abstract} 
Fault injection attacks represent a class of threats that can compromise embedded systems across multiple layers of abstraction, such as system software, instruction set architecture (ISA), microarchitecture, and physical implementation. 
Early detection of these vulnerabilities and understanding their root causes along with their propagation from the physical layer to the system software is critical to secure the cyberinfrastructure.

This present presents a comprehensive methodology for conducting \textit{controlled} fault injection attacks at the pre-silicon level and an analysis of the underlying system for root-causing behavior. As the driving application, we use the clock glitch attacks in AI/ML applications for critical misclassification. Our study aims to characterize and diagnose the impact of faults within the RISC-V instruction set and pipeline stages, while tracing fault propagation from the circuit level to the AI/ML application software. 
This analysis resulted in discovering a novel vulnerability through controlled clock glitch parameters, specifically targeting the RISC-V decode stage. 


\end{abstract}
\renewcommand\IEEEkeywordsname{Keywords}

\vspace{-0.75em}
\section{Introduction}\label{Introduction}
\vspace{-0.25em}
As embedded systems become increasingly integrated into critical infrastructures, ensuring their dependability and security is more essential than ever. Even minor disturbances in these systems can result in catastrophic failures or significant security breaches, where small disruptions can significantly impact results~\cite{FIA, Bilgiday,malik2020isolation,malik2021vrzycap,malik2022Obfuscation,malik2024enabling,malik2025epoch,karabulut2024themis}. 
In the context of AI/ML applications, a fault injection attack can cause a critical misclassification, which may have disastrous consequences, e.g., in autonomous vehicles, advanced weapons, and tactical systems.

Fault injection involves inducing malicious, transient faults through techniques such as clock glitches, electromagnetic interference, laser beams, and voltage manipulation, to trigger unintended behavior in the targeted system~\cite{bar2006sorcerer,Simplifi}. 
The key challenge in both attacking with and mitigating against fault injections is to understand how faults affect the physical hardware and how they propagate to the application layer, which involves analyzing circuit behavior, (micro)architecture, and system software.    
The advantages of doing this early at the pre-silicon level are obvious but it is even more challenging.


\section{Objectives}\label{Objectives}
\vspace{-0.25em}
This study is motivated by the need to improve modern microarchitectures' fault tolerance and develop robust tools for understanding and evaluating their vulnerability to fault injection attacks. 
As the case study, we focus on exploring the effects of clock glitch-based fault injections on AI/ML system behavior that can cause critical misclassifications. 
Our primary objective was to develop a framework that ranks and characterizes the potential vulnerability of a RISC-V code to fault injection attacks at the pre-silicon level. 
We specifically used the RISC-V based CV$32$e$40$x softcore processor to study, analyze, and instrument fault injection attacks. 
We consider the following specifications in our research.

\begin{figure*}[t]
\centering
    \vspace{-1em}
         \includegraphics[width =\textwidth]{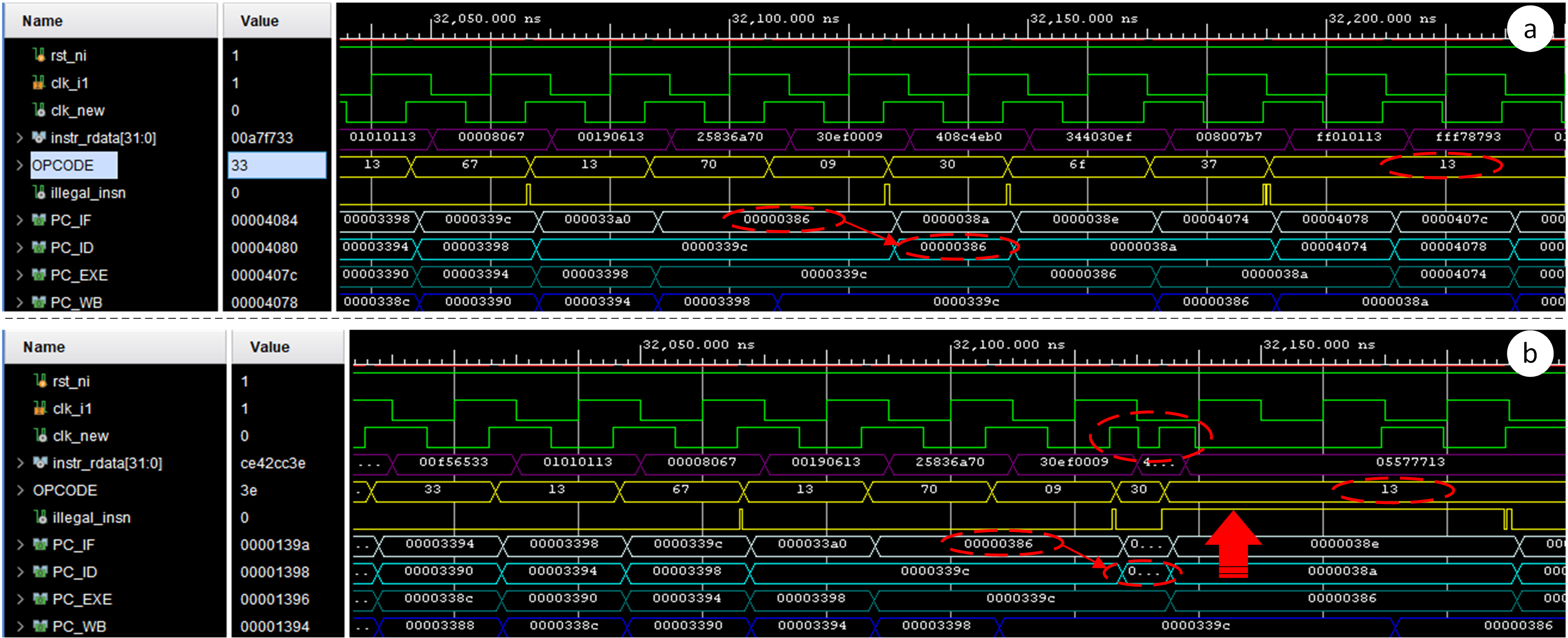}
\vspace{-2.25em}
	\caption{Pre-silicon risk assessment of RISC-V instructions (a) without and (b) with clock glitch in a binarized neural network inference design~\cite{eBNN}, a precise clock glitch converts the legal `load' instruction to an `illegal' instruction. This causes a misprediction by the neural network.}
 
\vspace{-1.75em}
	\label{fig: results}
\end{figure*}

\vspace{0.25em}
\begin{itemize}
    \item \textbf{Fast}: We develop a gate-level simulation framework as it is the earliest phase in the design that can generate reliable timing information.  This allows a faster simulations compared to, e.g., post place-and-route or SPICE-level simulations.

    \item \textbf{Precision}: The results are compared with the post-silicon hardware implementations and the simulation framework is calibrated to ensure correctness.

    \item \textbf{Comprehensive}: To make our findings more generic and span various hardware abstraction levels, we model the faults at the instruction set architecture level.

\end{itemize}

\section{Novelty}\label{Novelty}
\vspace{-0.25em}
We have performed root-cause analysis of various RISC-V instructions in an AI/ML software to identify and quantify points vulnerable to fault injection attacks within each processor pipeline stage. The key contributions of our research are:
\begin{enumerate}
    \item Quantifying the critical path timings of RISC-V instructions, providing insights into the reasons and the likelihood of successful attacks at each pipeline stage. 
    
    \item Discovering a novel RISC-V vulnerability at the decode stage, capable of turning a legal instruction into an illegal one.

    \item Conducting the analysis at the pre-silicon level, allowing sufficient time for the designers to identify and patch vulnerabilities before the final chip tape-out.
\end{enumerate}
We aim to provide a deeper understanding of guided fault attacks and their impact on modern microarchitectures.

\vspace{-0.5em}
\section{Proposed Methodolgy}\label{Prposed-Methodology}
\vspace{-0.25em}
Our pre-silicon fault injection attack analysis uses the RISC-V-based framework and Xilinx Vivado simulator to identify vulnerable instructions. The steps involved in our analysis are as follows:
\begin{enumerate}
    \item \textbf{Firmware.} RISC-V toolchain compiles the target source code, system files, and necessary dependencies.
     
     \item \textbf{Timing Analysis.} We analyze the design using Xilinx Vivado post-synthesis (gate-level) simulation to capture circuit-level timing information and potential violations.  
          
     \item \textbf{Risk Assessment Table (RAT).} RAT quantifies the critical path timings of RISC-V instructions for each pipeline stage to help identify the likelihood of attacks at each processor stage.
     
     \item \textbf{Fault Injection Attack.} Using RAT, we identify vulnerable points to introduce clock glitches to evaluate the system's robustness for RISC-V instructions at each pipeline stage. RAT enables precise targeting of RISC-V stages. For example, by targeting the decode stage failures, it identified a critical failure where legal instruction is misidentified as illegal.

     \item \textbf{Post-Silicon Validation.} Using RAT and the data collected at the pre-silicon level, we conduct a targeted FIA on the hardware implementation to validate our findings against post-silicon results.
\end{enumerate}

\begin{figure}[b]
\centering
\vspace{-2.0em}
         \includegraphics[scale =0.9]{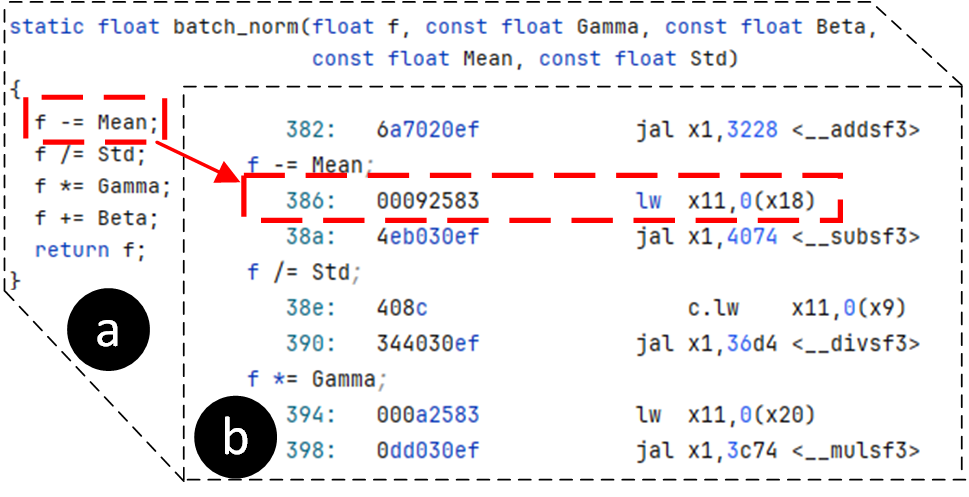}
\vspace{-2.0em}
	\caption{(a) Target C-code of binarized neural network inference and (b) the corresponding `load' instruction in the ISA subjected to clock-glitch attack.}
 
\vspace{-1.5em}
	\label{fig: code}
\end{figure}

\vspace{-0.5em}
\section{Results} \label{Results}
\vspace{-0.5em}
We evaluated using binarized neural network inference code at the pre-silicon level and simulating the design at the post-synthesis stage to ensure our findings closely align with post-silicon results~\cite{eBNN}. We analyzed RISC-V instructions for vulnerability to fault injection attacks by injecting clock glitches, resulting in an accuracy drop.

We discovered several clock glitch configurations that caused failures in the RISC-V decode stage, as shown in the bottom half of Figure~\ref{fig: results}. 
We derived that this failure is due to a timing violation at the \texttt{if\_id\_pipeline\_o} pipeline registers that synchronize the instruction and bus response in the RISC-V. 
As a result, when legal instructions such as load or store were processed, the clock glitch caused them to be misinterpreted as illegal, turning them into no-operation (NOP) instructions.  Fine-tuning this attack to target critical operations during inference, such as batch normalization or loop iteration checks, causes skipping them and changing the final classification result. Figure~\ref{fig: code}a displays the corresponding snippet of neural network C-code, while Figure~\ref{fig: code}b shows the corresponding assembly code and the attack targeting the `load' instruction. The results demonstrate the successful corruption of a legal `load' instruction into an `illegal' instruction. 





\vspace{-0.5em}
\section{Conclusions} \label{Conclusions}
\vspace{-0.25em}

As tactical systems and advanced defense platforms increasingly rely on embedded microcontrollers for mission-critical operations, a simulation framework that evaluates their susceptibility to fault injection attacks is required. 
These devices must be pre-silicon assessed to find and mitigate vulnerabilities early. 
We developed the first thorough framework to root-cause clock-glitch attacks at a pre-silicon level. 
Pre- and post-silicon testing showed our methodology can identify fault injection-vulnerable instructions. 
Our proposed solution can be used to build targeted defenses for identified vulnerabilities or to assess and rank different designs against such attacks.
\vspace{-.45em}

\vspace{-0.5em}
{\fontsize{12}{12}\selectfont{\bibliographystyle{IEEEtran}}
\bibliography{References}}

\begin{thebibliography}{10}
\providecommand{\url}[1]{#1}
\csname url@samestyle\endcsname
\providecommand{\newblock}{\relax}
\providecommand{\bibinfo}[2]{#2}
\providecommand{\BIBentrySTDinterwordspacing}{\spaceskip=0pt\relax}
\providecommand{\BIBentryALTinterwordstretchfactor}{4}
\providecommand{\BIBentryALTinterwordspacing}{\spaceskip=\fontdimen2\font plus
\BIBentryALTinterwordstretchfactor\fontdimen3\font minus \fontdimen4\font\relax}
\providecommand{\BIBforeignlanguage}[2]{{%
\expandafter\ifx\csname l@#1\endcsname\relax
\typeout{** WARNING: IEEEtran.bst: No hyphenation pattern has been}%
\typeout{** loaded for the language `#1'. Using the pattern for}%
\typeout{** the default language instead.}%
\else
\language=\csname l@#1\endcsname
\fi
#2}}
\providecommand{\BIBdecl}{\relax}
\BIBdecl

\bibitem{FIA}
A.~Barenghi, L.~Breveglieri, I.~Koren, and D.~Naccache, ``Fault injection attacks on cryptographic devices: Theory, practice, and countermeasures,'' \emph{Proceedings of the IEEE}, vol. 100, no.~11, pp. 3056--3076, 2012.

\bibitem{Bilgiday}
B.~Yuce, N.~F. Ghalaty, and P.~Schaumont, ``Improving fault attacks on embedded software using risc pipeline characterization,'' in \emph{Workshop on Fault Diagnosis and Tolerance in Cryptography}.\hskip 1em plus 0.5em minus 0.4em\relax IEEE, 2015, p.~97.

\bibitem{malik2020isolation}
A.~A. Malik, A.~Ullah, A.~Zahir, A.~Qamar, S.~K. Khattak, and P.~Reviriego, ``{Isolation design flow effectiveness evaluation methodology for Zynq SoCs},'' \emph{Electronics}, vol.~9, no.~5, p. 814, 2020.

\bibitem{malik2021vrzycap}
B.~Sultana, A.~Ullah, A.~A. Malik, A.~Zahir, P.~Reviriego, F.~B. Muslim, N.~Ullah, and W.~Ahmad, ``{VR-ZYCAP: A Versatile Resource-Level ICAP Controller for ZYNQ SOC},'' \emph{Electronics}, vol.~10, no.~8, p. 899, 2021.

\bibitem{malik2022Obfuscation}
N.~Nasir, A.~Ali~Malik, I.~Tahir, A.~Masood, and N.~Riaz, ``{Ephemeral Key-based Hybrid Hardware Obfuscation},'' in \emph{2022 19th International Bhurban Conference on Applied Sciences and Technology (IBCAST)}, 2022, pp. 646--652.

\bibitem{malik2024enabling}
A.~A. Malik, E.~Karabulut, A.~Awad, and A.~Aysu, ``{Enabling secure and efficient sharing of accelerators in expeditionary systems},'' \emph{Journal of Hardware and Systems Security}, vol.~8, no.~2, pp. 94--112, 2024.

\bibitem{malik2025epoch}
A.~A. Malik, E.~Karabulut, and A.~Aysu, ``{EPOCH: Enabling Preemption Operation for Context Saving in Heterogeneous FPGA Systems},'' \emph{arXiv preprint arXiv:2501.16205}, 2025.

\bibitem{karabulut2024themis}
\BIBentryALTinterwordspacing
E.~Karabulut, A.~A. Malik, A.~Awad, and A.~Aysu, ``{ THEMIS: Time, Heterogeneity, and Energy Minded Scheduling for Fair Multi-Tenant Use in FPGAs },'' \emph{IEEE Transactions on Computers}, no.~01, pp. 1--14, May 2025. [Online]. Available: \url{https://doi.ieeecomputersociety.org/10.1109/TC.2025.3566874}
\BIBentrySTDinterwordspacing

\bibitem{bar2006sorcerer}
H.~Bar-El, H.~Choukri, D.~Naccache, M.~Tunstall, and C.~Whelan, ``The sorcerer's apprentice guide to fault attacks,'' \emph{Proceedings of the IEEE}, vol.~94, no.~2, pp. 370--382, 2006.

\bibitem{Simplifi}
J.~Grycel and P.~Schaumont, ``Simplifi: hardware simulation of embedded software fault attacks,'' \emph{Cryptography}, vol.~5, no.~2, p.~15, 2021.

\bibitem{eBNN}
B.~McDanel, S.~Teerapittayanon, and H.~Kung, ``Embedded binarized neural networks,'' \emph{arXiv preprint arXiv:1709.02260}, 2017.

\end{thebibliography}

\end{document}